\documentclass[11pt]{article}
\usepackage{amsmath}
\usepackage{amssymb}
\usepackage{amsthm}
\usepackage{color}
\usepackage{setspace}
\usepackage[round]{natbib}
\usepackage{graphicx}
\usepackage{relsize,setspace}  
\usepackage{url}               

\usepackage{pgf,tikz}          
\usetikzlibrary{shapes,decorations,arrows,calc,arrows.meta,fit,positioning,shapes.arrows,shapes.geometric,shapes.multipart,decorations.pathmorphing,}

\newtheorem{theorem}{Theorem}

\newtheorem{definition}{Definition}

\newcommand{\E}{\mathbb{E}}

\newcommand{\ind}{\perp\!\!\!\!\perp} 

\usepackage{geometry}
\DeclareGraphicsExtensions{.pdf, .jpg, .png}
\setkeys{Gin}{width=0.85\textwidth}

\title{Time-varying confounding in epidemic intervention evaluations}

\author{Yichi Zhang$^1$ and Forrest W. Crawford$^{1,2,3,4,5}$  \\[1em] 
\normalsize 1. Department of Biostatistics, Yale School of Public Health\\
\normalsize 2. Department of Statistics \& Data Science, Yale University \\
\normalsize 3. Department of Ecology \& Evolutionary Biology, Yale University \\
\normalsize 4. Yale School of Management \\
\normalsize 5. RAND Corporation}

\date{\today}

\begin{document}

\maketitle

\begin{abstract}
\noindent Estimating the causal effect of a time-varying public health intervention on the course of an infectious disease epidemic is an important methodological challenge. During the COVID-19 pandemic, researchers attempted to estimate the effects of social distancing policies, stay-at-home orders, school closures, mask mandates, vaccination programs, and many other interventions on population-level infection outcomes. However, measuring the effect of these interventions is complicated by time-varying confounding: public health interventions are causal consequences of prior outcomes and interventions, as well as causes of future outcomes and interventions. Researchers have shown repeatedly that neglecting time-varying confounding for individual-level longitudinal interventions can result in profoundly biased estimates of causal effects. However, the issue with time-varying confounding bias has often been overlooked in population-level epidemic intervention evaluations. In this paper, we explain why associational modeling to estimate the effects of interventions on epidemic outcomes based on observations can be prone to time-varying confounding bias. Using causal reasoning and model-based simulation, we show how directional bias due to time-varying confounding arises in associational modeling and the misleading conclusions it induces. \\[1em]
\end{abstract}


\section{Introduction}

Over the past two decades, there has been an unprecedented rise in the occurrence of infectious disease outbreaks. These include the severe acute respiratory syndrome coronavirus (SARS-CoV) outbreak in 2003, the influenza A(H1N1) pandemic from 2009-2010, the Middle East respiratory syndrome coronavirus (MERS-CoV) outbreak in 2012, the West African Ebola virus disease epidemic from 2013-2016, the Zika virus epidemic from 2015-2016, and most recently, the coronavirus disease 2019 (COVID-19) pandemic \citep{jones2008global, smith2014global, torres2022global}. The COVID-19 pandemic alone has resulted in over 670 million cases and 6.8 million deaths globally \citep{NYTCOVID}. In light of the potential for future outbreaks of rapidly spreading pathogens, facilitated by global connectivity and climate change, a new era of infectious disease has been anticipated \citep{baker2022infectious}. 


In response to infectious threats, policymakers often enact public health interventions, such as stay-at-home orders, mask mandates, social distancing guidelines, and vaccination programs, to reduce the burden of infectious diseases, including infections, hospitalizations, and deaths in the population. Seeking to control the spread of infection, decision-makers act in response to past and current states of infection dynamics, as well as already-enacted interventions. Likewise, implemented interventions may alter the course of infection dynamics in the target population, as well as future intervention plans. For this reason, public health interventions are coupled to the past and future; in other words, they are endogenous to the dynamic infection process that policymakers seek to control. Approaches to estimating the effect of policymakers' interventions in this setting that merely condition on a sequence of time-varying interventions induce a form of bias called ``time-varying confounding'', in the causal estimand. 

In the related context of time-varying treatments to affect individual patient outcome trajectories, epidemiologists have established that failure to account for time-varying confounding can result in profoundly biased estimates of treatment effects \citep{mansournia2017handling, clare2019causal}. One standard approach for dealing with time-varying confounding is the g-formula \citep{robins1986a} which averages over time-varying confounders in a way that does not induce bias. Structural nested models (SNMs) \citep{robins1992g-estimation,robins1994correcting} and marginal structural models (MSMs) \citep{robins2000marginalA} are also widely used causal approaches proposed to estimate the effects of time-varying treatments. However, these methods are rarely used in the evaluation of population-level interventions in infectious disease epidemics. 

Despite the availability of causal methods for the evaluation of time-varying interventions, infectious disease researchers have almost exclusively relied on dynamic models of epidemic evolution, especially the deterministic SIR and SEIR classes of infectious disease transmission models \citep{kermack1927a,keeling2007modeling}, for estimation of intervention effects. Researchers typically start by either fitting an entire dynamic system to observed data or referring to scholarly literature for the estimation of model parameters. Next, infectious disease outcomes are generated under a counterfactual time-varying intervention that differs from the one that occurred in reality. The difference between the simulated counterfactual outcome and the observed outcome is then interpreted as the ``effect'' of the intervention, or sometimes interpreted as outcomes ``averted by the intervention'' \citep{chikina2020modeling, eikenberry2020to, gatto2020spread, lai2020effect, pei2020differential, prem2020the, radulescu2020management, zhang2020changes, rainisch2022estimated, shoukat2022lives, steele2022estimated, watson2022global}. 

Dynamic models, widely used by empirical researchers, are often considered structural and thus causal for modeling infectious disease evolutions and evaluating interventions. In theory, they can be considered a special case of the g-formula by specifying a certain class of disease data-generating processes. However, in practice, empirical researchers rarely discuss whether the compartments in the refined dynamic system models they propose form a sufficient set of time-varying confounders as a premise to help recover the causal effect of interest. This modeling approach becomes associational rather than causal when time-varying confounders are insufficiently collected, modeled, or improperly adjusted for. 

In this paper, we explain why using associational modeling practices to estimate counterfactual outcomes may result in time-varying confounding and biased estimates of causal estimands. We first describe a general infectious disease epidemic setting. We then outline an ideal causal estimand that captures the causal consequences of a static time-varying intervention under time-varying confounding. The correct identification formula for this target estimand is based on the g-formula, which properly adjusts for time-varying confounding. We then show analytically and via simulation that the statistical quantity recovered by any associational modeling approach is biased relative to the intended causal estimand. We propose conditions for a special set of static population-level epidemic interventions that lead to contradictory conclusions about their impact because of directional time-varying confounding bias. We conclude with advice for epidemiologists on making causal claims from epidemic models to support unbiased effect estimation.

\section{Setting}

Consider a discrete-time infection process consisting of outcomes, treatments, and covariates indexed by time $t$, $t = 1,...,T$. Let $Y_t$ be the observed infection-related outcome of interest (e.g., infections, hospitalizations, or deaths) at time $t$. Let $A_t$ be the population-level intervention (e.g., a stay-at-home order, mask mandate, or vaccination capacity) at time $t$. The intervention may be discrete- or continuous-valued, and can be implemented at any time point $t>0$. Let $X_t$ consist of time-varying covariates (e.g. diagnoses, hospitalizations, population mobility) at time $t$ other than the outcome of interest. We denote the observed history of outcomes, interventions, and covariates up to time $t$ as $\bar{Y}_{t}$, $\bar{A}_{t}$, and $\bar{X}_{t}$, respectively. 

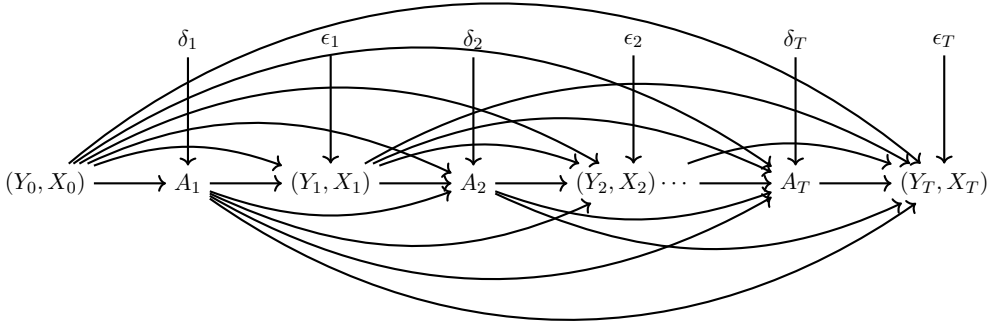
\begin{figure}
\scalebox{0.8}{
\centering
        \begin{tikzpicture}
        \node (Y_0_general) {$(Y_{0}, X_{0})$};
        \node[right = 1.2 of Y_0_general, text centered] (A_1_general) {$A_{1}$};
        \node[right = 1.2 of A_1_general, text centered] (Y_1_general) {$(Y_{1}, X_{1})$};
        \node[right = 1.2 of Y_1_general, text centered] (A_2_general) {$A_{2}$};
        \node[right = 1.2 of A_2_general, text centered] (Y_2_general) {$(Y_{2}, X_{2})\cdots$};
        \node[right = 1.2 of Y_2_general, text centered] (A_T_general) {$A_{T}$};
        \node[right = 1.2 of A_T_general, text centered] (Y_T_general) {$(Y_{T}, X_{T})$};

        \node[above = 1.8 of A_1_general, text centered] (delta_1_general){$\delta_{1}$};
        \node[above = 1.8 of Y_1_general, text centered] (epsilon_1_general){$\epsilon_{1}$};
        \node[above = 1.8 of A_2_general, text centered] (delta_2_general){$\delta_{2}$};
        \node[above = 1.8 of Y_2_general, text centered] (epsilon_2_general){$\epsilon_{2}$};
        \node[above = 1.8 of A_T_general, text centered] (delta_T_general){$\delta_{T}$};
        \node[above = 1.8 of Y_T_general, text centered] (epsilon_T_general){$\epsilon_{T}$};

        \draw[->, line width= 1] (Y_0_general) -- (A_1_general);
        \draw[->, line width= 1] (Y_0_general) to [out=20,in=160] (Y_1_general);
        \draw[->, line width= 1] (Y_0_general) to [out=25,in=155] (A_2_general); 
        \draw[->, line width= 1] (Y_0_general) to [out=30,in=150] (Y_2_general);
        \draw[->, line width= 1] (Y_0_general) to [out=35,in=145] (A_T_general);
        \draw[->, line width= 1] (Y_0_general) to [out=40,in=140] (Y_T_general);
        \draw[->, line width= 1] (A_1_general) --  (Y_1_general);
        \draw[->, line width= 1] (A_1_general) to [out=-20,in=-160] (A_2_general);
        \draw[->, line width= 1] (A_1_general) to [out=-25,in=-155] (Y_2_general);
        \draw[->, line width= 1] (A_1_general) to [out=-30,in=-150] (A_T_general);
        \draw[->, line width= 1] (A_1_general) to [out=-35,in=-145] (Y_T_general);
        \draw[->, line width= 1] (Y_1_general) -- (A_2_general);
        \draw[->, line width= 1] (Y_1_general) to [out=20,in=160] (Y_2_general);
        \draw[->, line width= 1] (Y_1_general) to [out=25,in=155] (A_T_general);
        \draw[->, line width= 1] (Y_1_general) to [out=30,in=150] (Y_T_general);
        \draw[->, line width= 1] (A_2_general) -- (Y_2_general);
        \draw[->, line width= 1] (A_2_general) to [out=-20,in=-160] (A_T_general);
        \draw[->, line width= 1] (A_2_general) to [out=-25,in=-155] (Y_T_general); 
        \draw[->, line width= 1] (Y_2_general) -- (A_T_general);
        \draw[->, line width= 1] (Y_2_general) to [out=20,in=160] (Y_T_general);
        \draw[->, line width= 1] (A_T_general) -- (Y_T_general);

        \draw[->, line width= 1] (delta_1_general) -- (A_1_general);
        \draw[->, line width= 1] (epsilon_1_general) -- (Y_1_general);
        \draw[->, line width= 1] (delta_2_general) -- (A_2_general);
        \draw[->, line width= 1] (epsilon_2_general) -- (Y_2_general);
        \draw[->, line width= 1] (delta_T_general) -- (A_T_general);
        \draw[->, line width= 1] (epsilon_T_general) -- (Y_T_general);
        
        \end{tikzpicture}}
    \caption{Causal directed acyclic graph (DAG) by the general DGP model. Arrows represent causal dependence of time-varying variables and interventions on the previous history and exogenous noises.} \label{fig:DAG_general}
\end{figure}

The stochastic data-generating process (DGP) evolves according to the following structural equations:
\begin{equation}\label{true_conditional_model}
\begin{split}
A_t|\{\bar{A}_{t-1}, \bar{Y}_{t-1}, \bar{X}_{t-1}\}  &= g_t(\bar{A}_{t-1}, \bar{Y}_{t-1}, \bar{X}_{t-1}, \delta_t)\\
(Y_t, X_t)|\{\bar{A}_t, \bar{Y}_{t-1}, \bar{X}_{t-1}\} &= f_t(\bar{A}_t, \bar{Y}_{t-1}, \bar{X}_{t-1}, \epsilon_t), 
\end{split}
\end{equation}
where $f_t(\cdot)$ and $g_t(\cdot)$ are arbitrary deterministic functions, and $\delta_t$ and $\epsilon_t$ are random variables. Figure \ref{fig:DAG_general} shows a causal directed acyclic graph (DAG) corresponding to this data-generating process. 

Let $Y_T^{\bar{a}_{T}}$ be the potential outcome \citep{rubin1978bayesian} at time $T$ under a counterfactual intervention history $\bar{a}_{T}$. In general, epidemiologists wish to learn about the average potential outcome $\E[Y_T^{\bar{a}_{T}}]$. The g-formula \citep{robins1986a} is one way of identifying this causal estimand: 
\begin{equation}\label{eq:g_general}
    \begin{split}
    \E[Y_T^{\bar{a}_T}] 
    = \idotsint &\E[Y_T|\bar{A}_T=\bar{a}_T, \bar{Y}_{T-1}=\bar{y}_{T-1}, \bar{X}_{T-1}=\bar{x}_{T-1}]\\
    &\prod_{t=1}^{T-1} p_t(y_{t}, x_{t}|\bar{a}_{t}, \bar{y}_{t-1}, \bar{x}_{t-1})\ \text{d}y_t\ \text{d}x_t.
    \end{split}
\end{equation}
The function $f_t$ and the distribution of $\epsilon_t$ in (\ref{true_conditional_model}) determine the conditional distribution $p_t(y_{t}, x_{t}|\bar{a}_{t}, \bar{y}_{t-1}, \bar{x}_{t-1})$ at each time $t$. The time-varying intermediate quantity $(Y_t,X_t)$ is called a \emph{time-varying confounder} if the effect of any intervention prior to time $t$ on any outcome after $t$ is mediated by it, and it is also a common cause of both treatments and outcomes subsequent to $t$. In other words, it is both the effect and cause of interventions at different time that might affect future outcomes. The g-formula recovers the expected potential outcome $Y_T^{\bar{a}_T}$ because it averages over the intermediate values of $(x_t,y_t)$ in a way that depends only on their conditional distribution, given the history of the process. 

\section{A general result on time-varying confounding bias for epidemic intervention evaluations}

In this section, we formally compare the causal estimand of interest, $\E[Y_T^{\bar{a}_{T}}]$, with its associational modeling approximation, $\E[Y_T|\bar{a}_T]$. We assume, generously, that empirical researchers do a perfect job of recovering the associational expectation $\E[Y_T|\bar{a}_T]$, allowing us to focus solely on the bias in identification rather than the error in estimation. For simplicity, and without loss of generality, we omit $X_t$, the time-varying confounders aside from the intermediate outcomes. The g-formula that identifies the true causal estimand is reduced to
\begin{equation}\label{eq:g_simple}
    \begin{split}
        \E[Y_T^{\bar{a}_{T}}] = \idotsint &\E[Y_T|\bar{A}_T=\bar{a}_T, \bar{Y}_{T-1}=\bar{y}_{T-1}]\prod_{t=1}^{T-1} p_t(y_{t}|\bar{a}_{t}, \bar{y}_{t-1}) \text{d}y_t ,
    \end{split}
\end{equation}
In comparison, any associational modeling approach approximates the average potential outcome above by the associational quantity
\begin{equation}\label{eq:obs_simple}
    \begin{split}
        \E[Y_T|\bar{a}_T] &= \idotsint \E[Y_T|\bar{A}_T=\bar{a}_T, \bar{Y}_{T-1}=\bar{y}_{T-1}]\prod_{t=1}^{T-1} p_t(y_{t}|\bar{a}_{T}, \bar{y}_{t-1}) \text{d}y_t \\
        &= \idotsint \E[Y_T|\bar{A}_T=\bar{a}_T, \bar{Y}_{T-1}=\bar{y}_{T-1}]\prod_{t=1}^{T-1} s_{t}(y_t; \bar{a}_T, \bar{y}_{t-1}) p_t(y_{t}|\bar{a}_{t}, \bar{y}_{t-1}) \text{d}y_t ,
    \end{split}
\end{equation}
where $s_{t}(y_t; \bar{a}_T, \bar{y}_{t-1}) = p(a_T, a_{T-1}, ..., a_{t+1}|\bar{a}_t, \bar{y}_{t})/p(a_T, a_{T-1}, ..., a_{t+1}|\bar{a}_t, \bar{y}_{t-1})$. The first equation is based on the definition of the associational quantity that conditions on $\bar{a}_T$, the entire history of time-varying interventions up to $T$. The second equation is by applying the Bayes' theorem:
\begin{equation*}
    p_t(y_{t}|\bar{a}_{T}, \bar{y}_{t-1}) 
    = \frac{p(a_T, a_{T-1}, ..., a_{t+1}|\bar{a}_t, \bar{y}_{t}) p_t(y_{t}|\bar{a}_{t}, \bar{y}_{t-1})}{p(a_T, a_{T-1}, ..., a_{t+1}|\bar{a}_t, \bar{y}_{t-1})}
    = s_{t}(y_t; \bar{a}_T, \bar{y}_{t-1})p_t(y_{t}|\bar{a}_{t}, \bar{y}_{t-1}).
\end{equation*}
This implies that the associational modeling approach is biased in recovering the correct causal estimand. The associational quantity employs an incorrect marginalizing distribution, which can be decomposed into the correct marginalizing distribution multiplied by some error term $s_{t}(y_t; \bar{a}_T, \bar{y}_{t-1})$. 

We make the following formal definition of time-varying confounding bias.
\begin{definition}[Time-varying confounding bias]
\label{defn:TVC}
For any static time-varying intervention of evaluation interest, $\bar{a}_T$, the \textbf{time-varying confounding bias} in identifying its average potential outcome is
\begin{equation*}
\E[Y_T|\bar{a}_T] - \E[Y_T^{\bar{a}_T}].
\end{equation*}
\end{definition}

The time-varying confounding bias is zero when $s_{t}(\bar{a}_T, \bar{y}_{t}) = 1, t = 1,..., T-1$ for any relevant $y_t$. When there is treatment endogeneity, updated information on an observed intermediate outcome, $y_t$, usually affects the probability of subsequent interventions. This possible change incurs a difference between the denominator of $s_{t}(\bar{a}_T, \bar{y}_{t})$, the probability of certain future interventions before observing $y_t$, and its numerator, that probability after observing $y_t$. Therefore, $s_t$ may not be equal to one when there is treatment endogeneity. We explain why time-varying confounding bias is nonzero and when it is directional by a formal theorem, presented below. Intuitively, three key components determine the time-varying confounding bias:
    \begin{enumerate}
        \item An endogenous dynamic rule of intervention assignment;
        \item A data-generating process of disease outcomes;
        \item A specific static intervention of evaluation interest.
    \end{enumerate}
We examine each of these three components in separate subsections to break down the issue of time-varying confounding bias in epidemic intervention evaluations. In the absence of a rule for preferential treatment assignment based on the history of disease evolution, time-varying confounding bias will not occur. In reality, the treatment rule is dynamic, though it may be either deterministic or random \citep{wen2021parametric}. The implemented policies interact with the disease transmission process over multiple time points, affecting the final outcome. Finally, even for the same rule of treatment assignment, there could be numerous possible realizations of different static time-varying interventions. The direction and magnitude of time-varying confounding bias are specific to and vary across different time-varying interventions. For a special class of interventions, we later propose and prove conditions under which directional time-varying confounding bias may lead to misleading causal evaluations.

\subsection{The dynamic rule of intervention assignment}

Confounding bias originates from an endogenous rule of intervention assignment. The propensity score is classically defined as the probability of a one-time intervention given all relevant confounders, which characterizes endogeneity \citep{rosenbaum1983central}. In the context of time-varying interventions, it becomes relevant to extend the definition of the propensity score to capture a series of probabilities of all remaining interventions at each intermediate time, conditioning on the history of the disease evolution. To formalize this extension, we define two types of prospective propensity scores within the true data-generating process.

\begin{definition}[Prospective propensity score]
\label{defn:PS}
Under the sequential ignorability assumption that the assignment of intervention does not depend on the potential outcomes given the observed history \citep{robins1986a},
\begin{equation*}
Y_T^{\bar{a}_{T}} \ind A_t | \bar{A}_{t-1}, \bar{Y}_{t-1},
\end{equation*}
for any history of interventions $\bar{a}_t = (a_0,\ldots,a_t)$ and outcomes $\bar{y}_t = (y_1,\ldots,y_{t})$, and any possible future interventions $(a_{t+1}^*,\ldots,a_T^*)$ of interest, the \textbf{lag-1 prospective propensity score} at $t = 1,..., T-1$ is
\begin{equation*}
p(A_T = a_T^*,\ldots A_{t+1} = a_{t+1}^*|\bar{A}_t = \bar{a}_t, \bar{Y}_{t-1} = \bar{y}_{t-1}) = p(a_T^*,\ldots a_{t+1}^*|\bar{a}_t, \bar{y}_{t-1}),
\end{equation*}
and the \textbf{lag-0 prospective propensity score} at $t = 1,..., T-1$ is
\begin{equation*}
p(A_T = a_T^*,\ldots A_{t+1} = a_{t+1}^*|\bar{A}_t = \bar{a}_t, \bar{Y}_{t} = \bar{y}_{t}) = p(a_T^*,\ldots a_{t+1}^*|\bar{a}_t, \bar{y}_{t}).
\end{equation*}
\end{definition}
In other words, the lag-1 prospective propensity score is the probability of all remaining interventions after $t$ based on the history until time $t-1$, and the lag-0 prospective propensity score is the probability of all remaining interventions after $t$ based on the history until time $t$. Once policymakers decide and implement the intervention $a_t$ at the start of time $t$, the realization and observation of the intermediate outcome $Y_t = y_t$ updates the probability of remaining interventions from the lag-1 to the lag-0 prospective propensity score. This latent ``update" on the probability of remaining interventions occurs due to policymakers' preferences for future intervention assignments based on the intermediate outcome at time $t$.

Therefore, the ratio between the lag-0 and lag-1 prospective propensity scores reveals the extent to which any observed $y_t$ skews the possibility of any series of remaining interventions being implemented. This ratio is exactly the difference between the g-formula and its associational approximation that we introduced in equation (\ref{eq:obs_simple}) as $s_{t}(y_t; \bar{a}_T, \bar{y}_{t-1})$. We formally define this ratio as a tool for investigating its impact on time-varying confounding bias.

\begin{definition}[Adaptive propensity score ratio]
\label{defn:ASR}
For any history of interventions $\bar{a}_t = (a_0,\ldots,a_t)$ and outcomes $\bar{y}_t = (y_1,\ldots,y_{t})$, and any future interventions $(a_{t+1}^*,\ldots,a_T^*)$ of interest, the \textbf{adaptive propensity score ratio} is the ratio between the lag-0 and lag-1 prospective propensity scores:
\begin{equation*}
s_t(y_t; a_T^*,\ldots,a_{t+1}^*, \bar{a}_t, \bar{y}_{t-1}) = \frac{p(a_T^*,\ldots a_{t+1}^*|\bar{a}_t, \bar{y}_{t})}{p(a_T^*, \ldots, a_{t+1}^*|\bar{a}_t, \bar{y}_{t-1})}, \quad t = 1,..., T-1.
\end{equation*}
\end{definition}
The adaptive propensity score ratio is greater than 1 when knowledge of $y_t$ increases the chance of implementing the future intervention of interest $(a_{t+1}^*,\ldots,a_T^*)$; it equals 1 when knowledge of $y_t$ does not alter the chance of its future implementation; it is less than 1 when knowledge of $y_t$ decreases the chance of its implementation.

\subsection{The DGP of disease outcomes}

As we move from the upstream cause of confounding, the endogenous rule of treatment assignment, to the downstream consequences of how it reweights intermediate outcomes, the next step in understanding the time-varying confounding bias in the final outcome is to investigate the connection between these intermediate outcomes and the final outcome. The mechanism of this connection is inherently determined by the true DGP of the disease evolution. Therefore, we use the previously defined adaptive propensity score to categorize the intermediate outcome state space based on reweighting patterns.

\begin{definition}[Upweighted/neutral/downweighted adaptations]
\label{defn:U/Dadpt}
For any history of interventions $\bar{a}_t = (a_0,\ldots,a_t)$ and outcomes $\bar{y}_t = (y_1,\ldots,y_{t})$, and any possible future interventions $(a_{t+1}^*,\ldots,a_T^*)$ of interest, the \textbf{upweighted adaptations} are
\begin{equation*}
\mathcal{U}_t(a_T^*,\ldots,a_{t+1}^*; \bar{a}_t, \bar{y}_{t-1}) = \{y_t: s_t(y_t, a_T^*,\ldots,a_{t+1}^*; \bar{a}_t, \bar{y}_{t-1}) > 1\},
\end{equation*}
the \textbf{neutral adaptations} are
\begin{equation*}
\mathcal{N}_t(a_T^*,\ldots,a_{t+1}^*; \bar{a}_t, \bar{y}_{t-1}) = \{y_t: s_t(y_t, a_T^*,\ldots,a_{t+1}^*; \bar{a}_t, \bar{y}_{t-1}) = 1\},
\end{equation*}
and the \textbf{downweighted adaptations} are
\begin{equation*}
\mathcal{L}_t(a_T^*,\ldots,a_{t+1}^*; \bar{a}_t, \bar{y}_{t-1}) = \{y_t: s_t(y_t, a_T^*,\ldots,a_{t+1}^*; \bar{a}_t, \bar{y}_{t-1}) < 1\}.
\end{equation*}
\end{definition}

Mathematically, the upweighted and downweighted adaptations are well-defined nonempty sets under the mild condition that the adaptive propensity score ratio is not almost surely constant with respect to the distribution of $Y_t$, given prior history.

The effect of these possible adaptations of the intermediate outcome at $t$ on the final outcome can be quantified through the following definition.
\begin{definition}[Moving marginal expectation]
\label{defn:MGE}
For any history of interventions $\bar{a}_t = (a_0,\ldots,a_t)$ and outcomes $\bar{y}_t = (y_1,\ldots,y_{t})$, and any possible future interventions $(a_{t+1}^*,\ldots,a_T^*)$ of interest, the \textbf{moving marginal expectation} is
\begin{equation*}
f_{T, t}(y_t, a_T^*,\ldots,a_{t+1}^*; \bar{a}_t, \bar{y}_{t-1}) 
= \idotsint y_T \prod_{s=t+1}^{T} p_s(y_{s}|a_s^*,\ldots,a_{t+1}^*, \bar{a}_t, \bar{y}_{s-1})\text{d}y_s.
\end{equation*}
\end{definition}
The moving marginal expectation captures the effect of the intermediate outcome $y_t$ on $y_T$, conditioned on a fixed history $(\bar{a}_t, \bar{y}_{t-1})$ and a specified sequence of remaining interventions $(a_{t+1}^*,\ldots,a_T^*)$. The expectation is ``marginal'' in that it integrates over all possible future outcomes after time $t$, given the history. It ``moves forward'' as $t$ increases, measuring the final outcome as a function of $y_t$ under the specified time-varying intervention.

\subsection{Directional bias in evaluating a special set of interventions}

The causal estimand and the time-varying confounding bias, as well as our previous definitions, depend on the specific time-varying intervention of evaluation interest. Naturally, the same rule of intervention assignment may yield different time-varying interventions due to randomness in disease evolution and probabilistic assignment of interventions. For certain time-varying interventions, the time-varying confounding bias is more severe than for others. Specifically, we focus on a particular set of static interventions that meet certain conditions. The conditions are based on an alignment between reweighting patterns caused by endogeneity and disease evolution patterns defined by the DGP. First, we formally define these conditions and interventions. We then detail the implications for inducing directional time-varying confounding bias.

\begin{definition}[Opportunistic intervention]
\label{defn:Oppint}
For any time-varying intervention $\bar{a}_T =\\ (a_0,\ldots,a_t, a_{t+1}^*,\ldots,a_T^*)$, it is \textbf{opportunistic} at $t$ if:\\ (i) for any $\bar{y}_{t-1} \in \{\bar{y}_{t-1}: p(a_T^*,\ldots a_{t+1}^*|\bar{a}_t, \bar{y}_{t-1}) > 0 \}$,
    \begin{equation*}
    \begin{split}
        & \inf_{y_t^L \in \mathcal{L}_t(a_T^*,\ldots,a_{t+1}^*; \bar{a}_t, \bar{y}_{t-1})} f_{T, t}(y_t^L, a_T^*,\ldots,a_{t+1}^*; \bar{a}_t, \bar{y}_{t-1}) \\
        \geq & \sup_{y_t^U \in \mathcal{U}_t(a_T^*,\ldots,a_{t+1}^*; \bar{a}_t, \bar{y}_{t-1})} f_{T, t}(y_t^U, a_T^*,\ldots,a_{t+1}^*; \bar{a}_t, \bar{y}_{t-1});
    \end{split}
    \end{equation*}
and (ii) there exists a set of adaptations $\mathcal{A}_t \subseteq (\mathcal{L}_t \cup \mathcal{U}_t)(a_T^*,\ldots,a_{t+1}^*; \bar{a}_t, \bar{y}_{t-1})$ and some $m \geq 0$ such that
    \begin{equation*}
        \int_{\mathcal{A}_t} |1 - s_t(y_t; \bar{a}_T, \bar{y}_{t-1})|p_t(y_t|\bar{a}_t, \bar{y}_{t-1}) \text{d}y_t > 0,
    \end{equation*}
and
    \begin{equation*}
    \begin{split}
        &\inf_{y_t^A \in \mathcal{A}_t} |f_{T, t}(y_t^A, a_T^*,\ldots,a_{t+1}^*; \bar{a}_t, \bar{y}_{t-1}) - \inf_{y_t^L \in \mathcal{L}_t} f_{T, t}(y_t^L, a_T^*,\ldots,a_{t+1}^*; \bar{a}_t, \bar{y}_{t-1})| \geq m.
    \end{split}
    \end{equation*}
\end{definition}
In other words, condition (i) states that, for an opportunistic intervention, the upweighted adaptations of intermediate outcomes that make the implementation for the remaining interventions more likely will also result in fewer adverse outcomes than the downweighted adaptations do. Condition (ii) introduces the weak assumption that the difference in outcomes produced by the two sets of adaptations is nonzero on at least a subset where the adaptive propensity score ratio deviates from one, as measured by the integral of their absolute distance.

The following result formalizes the notion of directional bias as a consequence of an opportunistic time-varying intervention. The proof is provided in the Appendix.
\begin{theorem}[Time-varying confounding bias from an opportunistic intervention]
\label{thrm:TVCB}
For a time-varying intervention $\bar{a}_T$ of evaluation interest, \textbf{the time-varying confounding bias is negative} if the intervention is \textbf{opportunistic} at every (and at least one) $t$ such that the adaptive propensity score ratio $s_t(y_t, a_T,\ldots,a_{t+1}; \bar{a}_t, \bar{y}_{t-1})$ is not almost surely constant.
\end{theorem}
The causal estimand of effect of an adverse outcome under an opportunistic time-varying intervention is thus underestimated by the associational approximation.

Given that researchers often consider no-intervention as the counterfactual control for assessing the effect of a factual intervention, an overly optimistic evaluation of disease transmission under no-intervention may lead to serious consequences, such as the under-promotion of effective, life-saving measures. In the next section, we present a realistic example in which no-intervention is opportunistic.

\section{Example: Directional bias under opportunistic no-intervention}

We describe a simple result that follows from Theorem \ref{thrm:TVCB}, where the counterfactual no-intervention scenario is opportunistic, resulting in directional bias. We introduce the following definitions for a general setup of a DGP for the disease outcome process and a rule for intervention assignment.

\begin{definition}[Monotonically increasing disease outcome process]
\label{defn:MDOP}
For any disease data generating process, it is a \textbf{monotonically increasing disease outcome process} if the moving marginal expectations for any $T > 2$ are monotonically increasing functions with respect to $y_t$ at each $t = 1 ,..., T-1$.
\end{definition}

For example, when the outcome of interest is the number of cumulative infections, hospitalizations, or deaths, the monotonically increasing disease outcome process assumes that the number of final cumulative infections increases if there is an increase in intermediate cumulative infections.

\begin{definition}[Self-consistent rule of intervention assignment]
\label{defn:SC}
For any time-varying intervention $\bar{a}_T$, it is realized under a \textbf{self-consistent rule of intervention assignment} if
\begin{equation*}
    p(A_{t+1} = 1|\bar{A}_{t}, \bar{Y}_{t}) = 
      1 \text{ if $\sum_{s=1}^{t} A_s = 0$ and $Y_t > \tilde{y}$}
\end{equation*}
\end{definition}

In other words, the intervention starts when a certain threshold for the intermediate outcome is reached. After this point, no specific restrictions are imposed on rules for stopping or restarting the intervention. Therefore, the self-consistent rule is dynamic and may be either deterministic or random, depending on the remaining part of the rule \citep{wen2021parametric}. We use $\tilde{y}$ to denote the threshold at which the intervention first begins. Combining the above definitions, the self-consistent rule of intervention assignment specifies that the intervention begins if cumulative outcomes exceed a certain level. This part of the rule is deterministic based on observed intermediate outcomes and interventions, but it remains probabilistic in terms of randomness before observing the intermediate outcome within the stochastic disease outcome process.

Suppose that, at the end of the infection process, policymakers aim to evaluate a counterfactual control scenario of no intervention using an associational model, $\E[Y_T|\bar{a}_T]$. Let $\bar{a}_T = (a_1, \ldots, a_T) = (0, \ldots, 0)$. We check its opportunistic property, relative to the monotonically increasing disease outcome process and the self-consistent rule of intervention assignment, at each $t$ in the Appendix. By Theorem \ref{thrm:TVCB}, the time-varying confounding bias is negative.

Intuitively, in the counterfactual scenario of no intervention, at the beginning of the disease process, each time where the intervention does not start is due to the observed cumulative outcomes being lower than a threshold. This process sequentially filters out those counterfactual trajectories of infection processes with more early outcomes, leading to underestimation of the outcome by the associational model. This is intuitive, as only in rare cases would the disease progress so slowly that the policy would never be implemented due to the threshold not being met. Interpreting the associational outcome as if it were a causal outcome under the counterfactual time-varying intervention, had it been exogenously implemented, would be highly misleading.

\section{Simulation}

We conduct a simulation to verify and demonstrate directional bias in the context of the opportunistic no-intervention example above. The setup requires a DGP for the disease evolution process and an endogenous rule of intervention assignment. First, for the true disease DGP, we consider a stochastic discrete-time Susceptible-Infected-Recovered (SIR) model \citep{allen2017primer}. Specifically, consider a population of $N$ individuals whose members are either susceptible to infection, infected, or recovered and unable to be infected or infect others. Let $S_t$, $I_t$, and $R_t$ be the number of susceptible, infected, and recovered individuals at time $t$, respectively. The outcome of interest is the cumulative infections (or the proportion of cumulative infections within the total population) up to time $T$, defined as $Y_T = N - S_T$ (or $Y_T = 1 - S_T/N$). The time-varying intervention sequence $\bar{A}_T$ interacts with the disease process at each time $t$, affecting its instant and future progression. The recurrent model for the DGP is
\begin{equation}\label{eq:SIR_SDE}
\begin{pmatrix}
S_{t}\\
I_{t}\\
R_{t}
\end{pmatrix}
    =
\begin{pmatrix}
S_{t-1} + \text{d}S_{t}\\
I_{t-1} + \text{d}I_{t}\\
R_{t-1} + \text{d}R_{t}
\end{pmatrix}
    =
\begin{pmatrix}
S_{t-1} - \frac{e^{\lambda A_t}}{N} \beta S_{t-1} I_{t-1} - \epsilon_{1t}\\
I_{t-1} + \frac{e^{\lambda A_t}}{N} \beta S_{t-1} I_{t-1} - \gamma I_{t-1}  + \epsilon_{1t} - \epsilon_{2t}\\
R_{t-1} + \gamma I_{t-1} + \epsilon_{2t}
\end{pmatrix}.
\end{equation}
We detail the choice of parameters and the distributions of the stochastic terms as follows: 
\begin{enumerate}
\item the size of the closed population is set to be $N = 1,000,000$; 
\item the initial numbers of susceptible, infected, and recovered individuals are set to be $S_0 = N - I_0$, $I_0 = 200$, and $R_0 = 0$; 
\item the duration for the simulation and the terminal point are $T = 100$ days;
\item the average duration of infection, $1/\gamma$, is fixed to be $7$ days; 
\item the contact rate is set to be $\beta = 2/7$ to return a basic reproduction number of $\mathcal{R}_0 = 2$; 
\item the strength of the intervention is set to be $\lambda=-0.2$ which scales the contact rate by a factor of $e^{-0.2} \approx 0.8$;
\item the random variations in new infections and new recoveries, denoted by $\epsilon_{1t}$ and $\epsilon_{2t}$, follow independent truncated normal distributions, $\mathcal{N}(0, 500\frac{e^{\lambda A_t}}{N} \beta S_{t-1} I_{t-1})$ truncated on $[-\frac{e^{\lambda A_t}}{N} \beta S_{t-1} I_{t-1}, S_{t-1} - \frac{e^{\lambda A_t}}{N} \beta S_{t-1} I_{t-1}]$, and $\mathcal{N}(0, 500\gamma I_{t-1})$ truncated on $[-\gamma I_{t-1}, N - R_{t-1} - \gamma I_{t-1}]$. In other words, the variances in these normal distributions are proportional to fixed increments in new infections and recoveries, scaled by an overdispersion parameter of 500. The truncation intervals ensure that new infections and recoveries are nonnegative, and that the sizes of updated compartments do not exceed $N$.
\end{enumerate}

\begin{figure}
    \centering
    \includegraphics[scale=0.23]{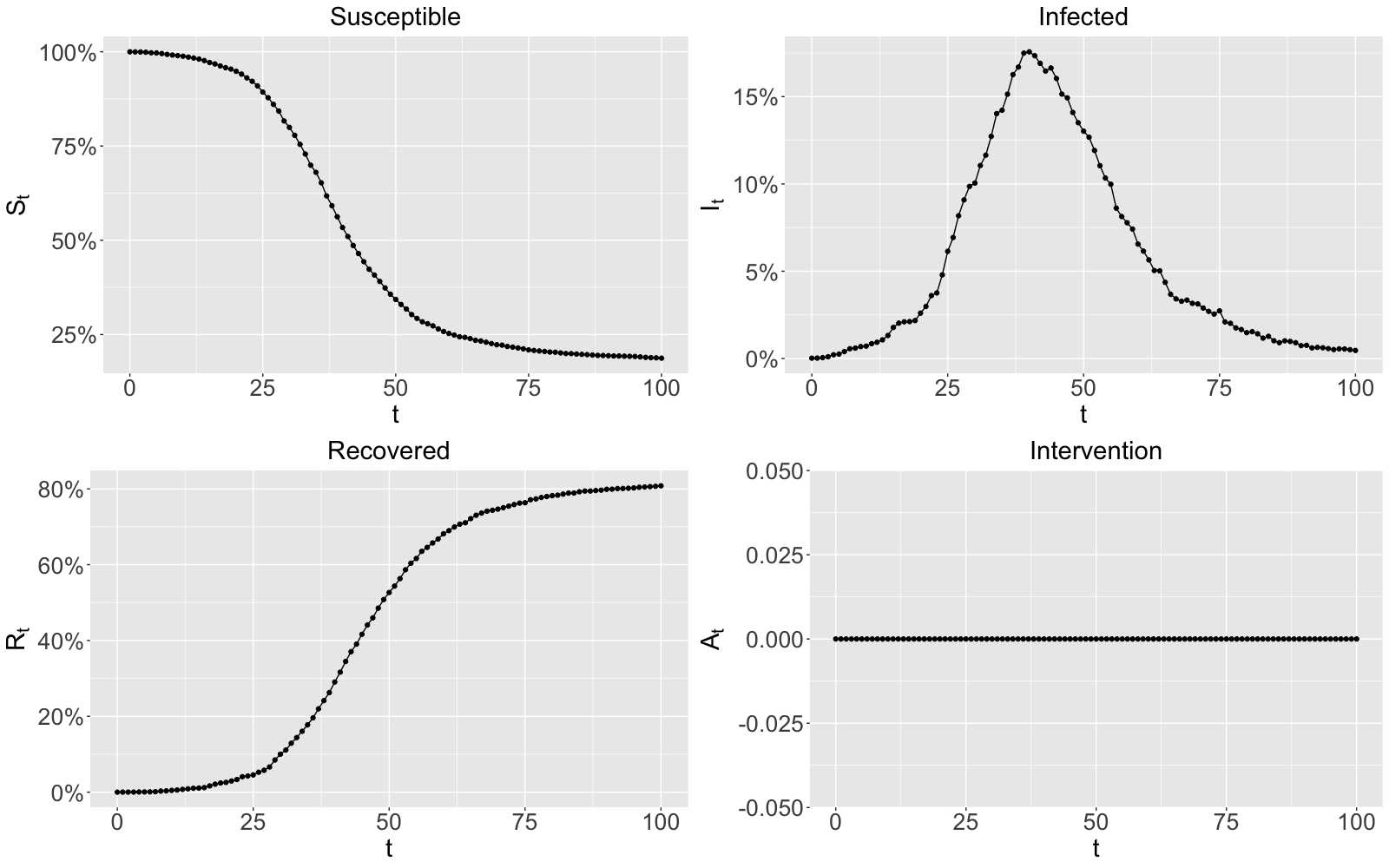}
    \caption{A single random realization of the trajectories of disease compartments under no intervention from the stochastic discrete-time SIR model.}\label{pic:SIR_sample}
\end{figure}

To compute the causal estimand under no-intervention, $\E[Y_T^{\bar{a}_T = (0, \ldots, 0)}]$, we simulate the above process $100,000$ times and calculate the average outcome, $Y_T = 1 - S_T/N$. We take a random sample of the disease evolution process as a brief visual illustration of the setup of the disease evolution in Fig. \ref{pic:SIR_sample}.


To simulate biased associational outcomes under no intervention, we impose self-consistent rules of intervention assignment (Definition \ref{defn:SC}) by specifying $\tilde{y}$, the intervention threshold of cumulative outcomes. We vary the intervention threshold $\tilde{y}$ to be $5\%$, $10\%$, $15\%$, $20\%$, $25\%$, $30\%$ of the population. For each $\tilde{y}$, we calculate the biased associational quantities by simulating the SIR disease process $100,000$ times.


We aim to show: 
\begin{enumerate}

\item The consequence: the time-varying confounding bias evolves over time and accumulates toward the negative direction as $t$ increases under the opportunistic no-intervention scenario.

\item The rationale: the time-varying confounding bias at each $t < T$ depends on both past and future interventions of interest. The history of no intervention is fixed, affecting the disease dynamics at $t$. The future of no intervention has not yet been realized but is made more likely by certain outcome states, governed by the dynamic rule of intervention. Therefore, the evolution of time-varying confounding bias is expected to depend on the intervention threshold $\tilde{y}$ even when the intervention is never implemented (this is the no-intervention treatment of interest).

\end{enumerate}

\begin{figure}
    \centering
    \includegraphics[scale=0.23]{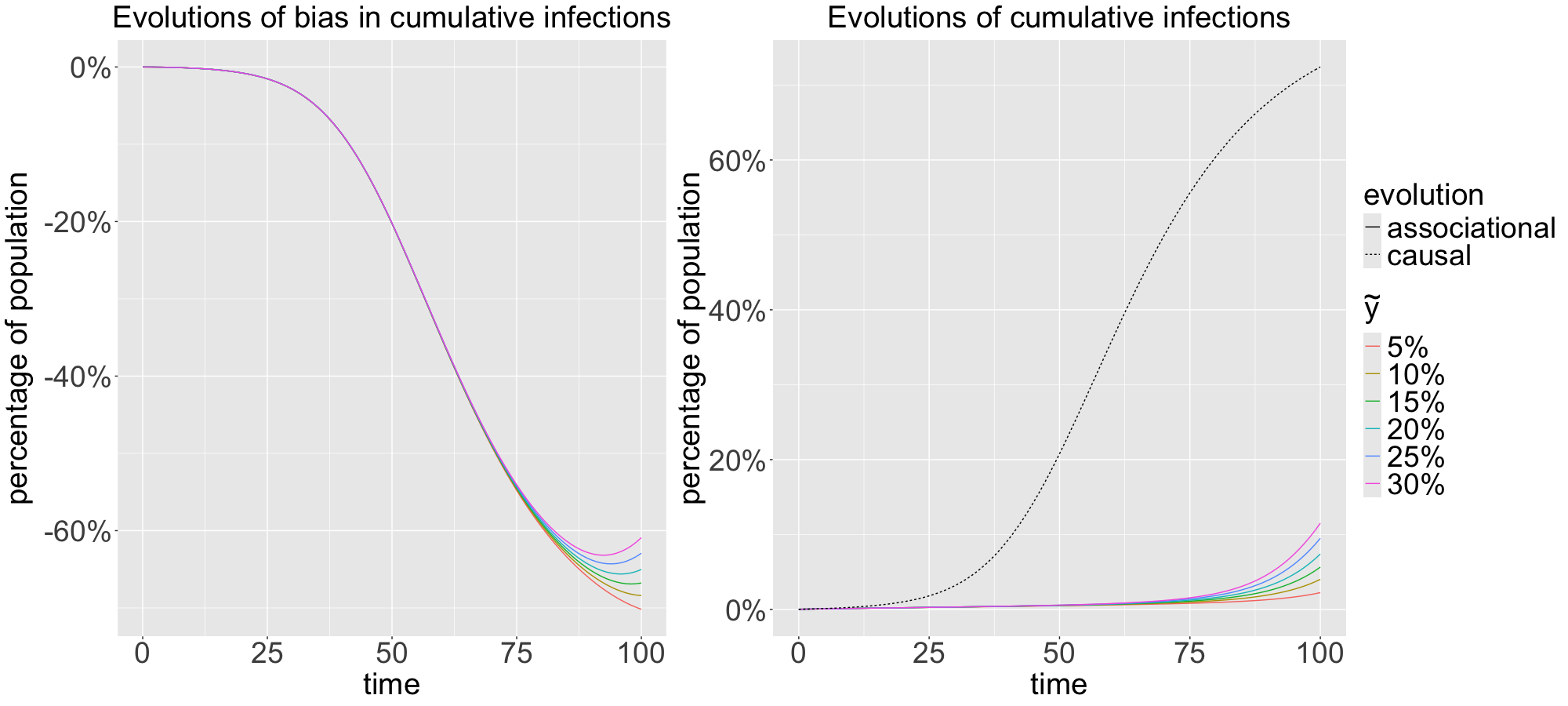}
    \caption{Left: evolutions of time-varying confounding bias $\E[Y_t|\bar{a}_T] - \E[Y_t^{\bar{a}_T}]$ under no intervention, i.e. $\bar{a}_T = (0, \ldots, 0)$ (x-axis: time $t$, y-axis: the time-varying confounding bias); Right: evolution of cumulative infections by the correct causal quantity $\E[Y_t^{\bar{a}_t}]$ (dashed line) vs. the incorrect associational approximations $\E[Y_t|\bar{a}_T]$ (solid lines) under no intervention (x-axis: time $t$, y-axis: the outcome of cumulative infections). The intervention threshold $\tilde{y}$ (percentage of population) is never reached for no-intervention cases; however, it still affects $\E[Y_T|\bar{a}_T = (0, \ldots, 0)]$ through the disease dynamics associated with no intervention under the corresponding rule of intervention assignment.}\label{pic:TV_cfdn_evolution}
\end{figure}

\begin{figure}
    \centering
    \includegraphics[scale=0.23]{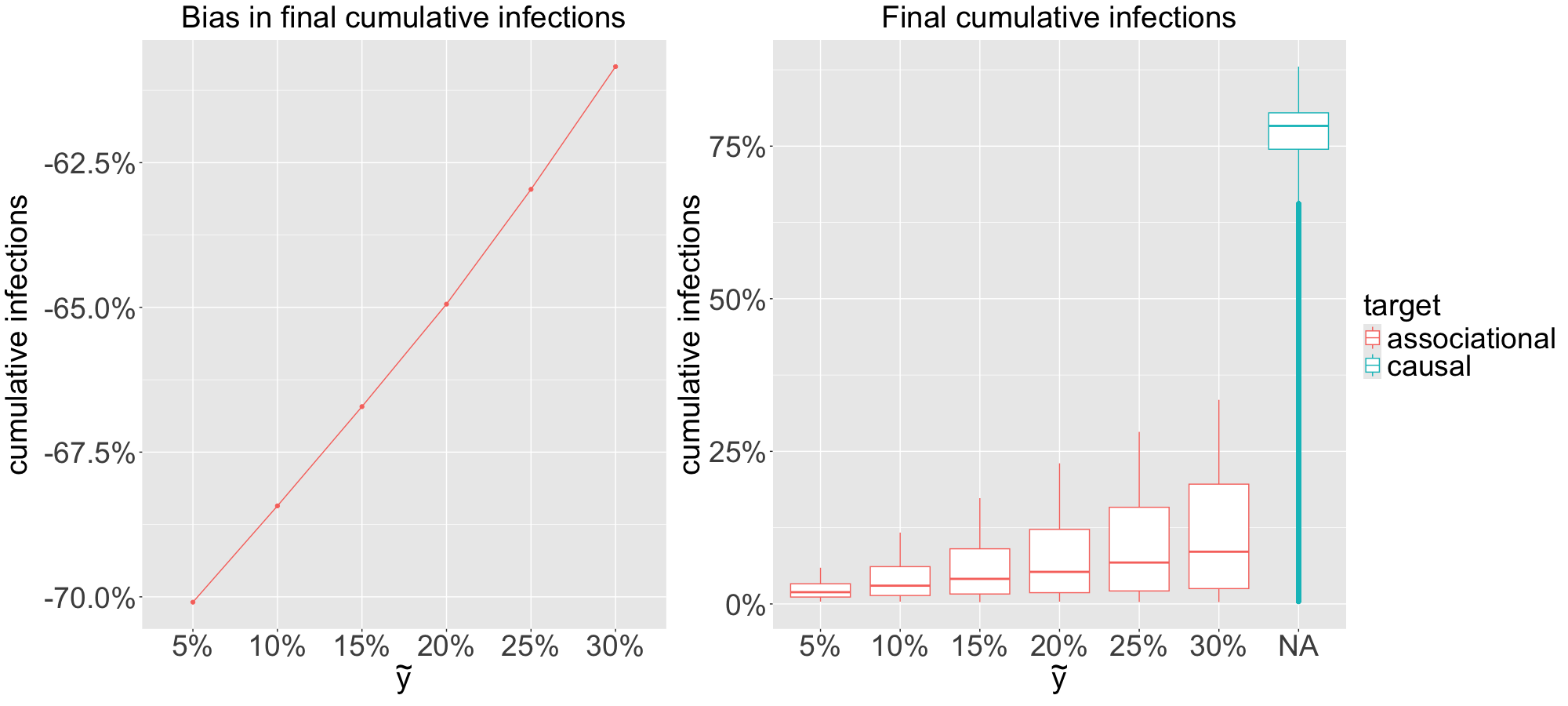}
    \caption{Left: time-varying confounding bias $\E[Y_T|\bar{a}_T] - \E[Y_T^{\bar{a}_T}]$ under no intervention, i.e. $\bar{a}_T = (0, \ldots, 0)$ (x-axis: the intervention threshold $\tilde{y}$ (percentage of population), exceeds which the intervention will start, y-axis: the time-varying confounding bias); Right: the correct causal estimand $\E[Y_T^{\bar{a}_T}]$ (blue) vs. the incorrect associational approximations $\E[Y_T|\bar{a}_T]$ (red) under no intervention (x-axis: the intervention threshold $\tilde{y}$, y-axis: the outcome of cumulative infections). The intervention threshold $\tilde{y}$ is never reached for no-intervention cases; however, it still affects $\E[Y_T|\bar{a}_T = (0, \ldots, 0)]$ through the disease dynamics associated with no intervention under the corresponding rule of intervention assignment.}\label{pic:TV_cfdn_bias}
\end{figure}

The results are presented in Figure \ref{pic:TV_cfdn_evolution} and Figure \ref{pic:TV_cfdn_bias}. The entire cumulative process of time-varying confounding bias across $t = 0,\ldots, T$ is shown in Figure \ref{pic:TV_cfdn_evolution}. On its left side, the plot shows that time-varying confounding bias by the associational quantities quickly accrues in a negative direction. This is the direct consequence of an opportunistic no-intervention. The multiple evolutionary trajectories of bias are under different values of $\tilde{y}$, which controls when the intervention starts in the corresponding rule of intervention assignment. As expected, even though the intervention never begins under this no-intervention scenario, the evolutionary patterns of bias still vary with the choice of intervention threshold $\tilde{y}$. 

The right-side plot in \ref{pic:TV_cfdn_evolution} decomposes the trend described above. It juxtaposes the natural evolution of cumulative infections by the correct causal approach with the biased associational quantity under different rules governed by $\tilde{y}$. The former rises sharply under no intervention but the latter stays low, which increases the probability that the intervention never starts during the process. The heterogeneity in evolutionary patterns by $\tilde{y}$ starts to exhibit at the end of the process, when a moderate increase in cumulative infections towards $\tilde{y}$ on average does not escalate the ``risk" of starting the intervention before $t = T$.

Figure \ref{pic:TV_cfdn_bias} shows the results of time-varying confounding at a single terminal time point $t = T$. The plot on the left side shows that the time-varying confounding bias ranges between $-60\%$ and $-70\%$ of the entire population. Specifically, the plot on the right side shows that the true number of cumulative infections under no intervention, according to the causal estimand, is expected to exceed $75\%$ of the population. In contrast, the associational outcomes conclude that the number is no more than $10\%$. Therefore, the associational outcome significantly underestimates cumulative infections in the absence of any intervention to control disease transmission.

As the threshold $\tilde{y}$ decreases from $30\%$ to $5\%$, the magnitude of the time-varying confounding bias increases. This can be explained by the fact that no-intervention cases under a lower $\tilde{y}$ are associated with slower disease evolutions and lower infection peaks. This results in fewer outcomes, which leads to more serious bias by the associational outcome.

In summary, the simulation verifies that the time-varying confounding bias is negative under an opportunistic no-intervention. It accumulates during the disease evolution process. The magnitude of bias and its evolutionary pattern depend on the endogenous dynamic rule of intervention. Despite conditioning on a counterfactual scenario of no intervention, the intervention threshold of cumulative infections, $\tilde{y}$, influences the evolutionary patterns of the associational quantity. The consequence of time-varying confounding bias over $60\%$ of the population may result in overly optimistic evaluations of no intervention.

\section{Discussion}

This work, while conceptual in nature, explains why common real-world practices in infectious disease modeling are prone to bias when estimating the causal effects of epidemic interventions. Although time-varying confounding has been recognized as a crucial issue in causal inference for decades \citep{robins1986a} and explained more plainly in recent years \citep{mansournia2017handling}, the general awareness of its impact on causal conclusions is still lacking, especially in population-level infectious disease research. Connecting popular methods that support decision-making with the issue of bias arising from time-varying confounding highlights the need to understand this issue to avoid harmful consequences, such as promoting ineffective interventions, or under-using effective ones.

We support this conclusion through intuition, theory, and simulation. Intuitively, decisions by policymakers to implement interventions are governed by latent rules, affected by past observations and potentially impacting future outcomes. This requires consideration of factors that contribute to decisions and interact with disease dynamics. However, discussions on constructing dynamic models rarely address adjustments for these variables or include models that account for intervention endogeneity. Without these adjustments, modeling approaches often collapse into associational analysis.

We demonstrate the fundamental structural difference between a causal estimand and an associational outcome. Focusing on a special set of opportunistic interventions, we dissect the issue of time-varying confounding and formalize the intuition behind extreme directional bias. This applies to realistic cases of counterfactual evaluations under no intervention and general disease outcome processes. We verify through simulations that time-varying confounding bias can be over the majority of the population. This bias can lead to disastrous decisions such as implementing inadequate measures to prevent disease transmission.


Our framework provides a theoretical decomposition of time-varying confounding bias rather than a direct simple solution. The adaptive propensity score ratio, used as a weight to analyze bias by the associational outcome, cannot be readily calculated without knowledge of the disease outcome process. Calculating its inverse for recovering the causal estimand from an associational quantity requires knowledge of both the disease DGP and the dynamic rule of intervention assignment. Many causal inference methods require less information to identify the causal estimand. For example, the marginal structural model (MSM) \citep{robins2000marginalA} requires a marginal outcome model and stabilized inverse probability weights to identify and estimate the causal estimand, without requiring models for the entire DGP. Recently, \citet{bonvini2022causal} derived a simplistic marginal structural model based on classic dynamic models in infectious disease. They adjusted for time-varying confounders by the inverse probability weights and conducted a sensitivity analysis for their choices of confounders in an empirical study.

Time-varying confounding bias can also be understood by comparing a dynamic treatment rule with a static time-varying intervention \citep{wen2021parametric}. Associational quantities like observed outcomes can be sensibly interpreted as causal quantities under the implemented dynamic rule of treatment assignment but not under the observed static time-varying intervention. Although the observed static intervention arises from the implemented dynamic rule, conditioning on it as if it is exogenous without considering its endogenous origin removes context, inducing time-varying confounding bias in associational outcomes.

Addressing time-varying confounding bias in infectious disease necessitates the development and deployment of causal inference methodology tailored to disease processes, including existing methods that deal with the time-varying confounding \citep{robins1986a, robins1992g-estimation, robins1994correcting, robins1997estimation, robins2000marginalA, bang2005doubly, bonvini2022causal}. These approaches require collecting and adjusting for time-varying confounders, either by modeling them as compartments or using them to estimate the dynamic rule of intervention. Such adjustments are essential for addressing time-varying confounding bias in causal epidemic evaluations.

\appendix 

\section{Appendix}

\subsection*{Proof for Theorem \ref{thrm:TVCB}}

\begin{proof}

The moving marginal estimands under $\bar{a}_T$ are defined as
\begin{equation*}
f_{T, t}(y_t, a_T,\ldots,a_{t+1}; \bar{a}_t, \bar{y}_{t-1}) 
= \idotsint y_T \prod_{l=t+1}^{T} p_l(y_{l}|\bar{a}_l, \bar{y}_{l-1})\text{d}y_l.
\end{equation*}

The causal estimand can be expressed in terms of the marginal moving estimands as
\begin{equation}
    \begin{split}
        \E[Y_T^{\bar{a}_T}] 
        &= \idotsint \E[Y_T|\bar{A}_T=\bar{a}_T, \bar{Y}_{T-1}=\bar{y}_{T-1}]\prod_{l=1}^{T-1} p_l(y_{l}|\bar{a}_{l}, \bar{y}_{l-1}) \text{d}y_l \\
        &= \idotsint f_{T, {T-1}}(y_{T-1}, a_T; \bar{a}_{T-1}, \bar{y}_{T-2}) \prod_{l=1}^{T-1} p_l(y_{l}|\bar{a}_{l}, \bar{y}_{l-1}) \text{d}y_l \\
        &= \idotsint f_{T, t}(y_t, a_T,\ldots,a_{t+1}; \bar{a}_t, \bar{y}_{t-1}) \prod_{l=1}^{t} p_l(y_{l}|\bar{a}_{l}, \bar{y}_{l-1}) \text{d}y_l \\
        &= \int f_{T, 1}(y_1, a_T,\ldots,a_{2}; a_1, y_0) p_1(y_1|a_{1}, y_0)\text{d}y_1.
    \end{split}
\end{equation}

For any $t = 1,..., T-1$ and any $\bar{y}_{t-1} \in \{\bar{y}_{t-1}: p(a_T,\ldots a_{t+1}|\bar{a}_t, \bar{y}_{t-1}) > 0 \}$, notice that
\begin{equation}
\label{prop:eq1}
\begin{split}
&\hspace{0.5em} \int (s_t(y_t; \bar{a}_T, \bar{y}_{t-1}) - 1) p_t(y_t|\bar{a}_t, \bar{y}_{t-1}) \text{d}y_t \\
= &\hspace{0.5em} \frac{1}{p(a_T, \ldots, a_{t+1}|\bar{a}_t, \bar{y}_{t-1})}\int p(a_T,\ldots a_{t+1}|\bar{a}_t, \bar{y}_{t}) p_t(y_t|\bar{a}_t, \bar{y}_{t-1}) \text{d}y_t - 1 \\
= &\hspace{0.5em} 0.
\end{split}
\end{equation}
Therefore, for any $p(a_T, a_{T-1}, ..., a_{t+1}|\bar{a}_t, \bar{y}_{t})$ not almost surely constant with regard to $p(y_t|\bar{a}_t, \bar{y}_{t-1})$, $\mathcal{L}_t(a_T,\ldots,a_{t+1}; \bar{a}_t, \bar{y}_{t-1})$ and $\mathcal{U}_t(a_T,\ldots,a_{t+1}; \bar{a}_t, \bar{y}_{t-1})$ are well-defined non-empty sets with nonzero probability conditioning on $\bar{a}_t$ and $\bar{y}_{t-1}$.

Suppose we have an opportunistic intervention. We denote 
\begin{equation*}
    b_t(\bar{a}_T, \bar{y}_{t-1}, \bar{x}_{t-1}) = \inf_{y_t^L \in \mathcal{L}_t(a_T,\ldots,a_{t+1}; \bar{a}_t, \bar{y}_{t-1})} f_{T, t}(y_t^L, a_T,\ldots,a_{t+1}; \bar{a}_t, \bar{y}_{t-1}),
\end{equation*}
then we have
\begin{equation}
\label{prop:eq2}
\begin{split}
&\hspace{0.5em} \int (s_t(y_t; \bar{a}_T, \bar{y}_{t-1})-1)(f_{T, t}(y_t, a_T,\ldots,a_{t+1}; \bar{a}_t, \bar{y}_{t-1})-b_t(\bar{a}_T, \bar{y}_{t-1})) p_t(y_t|\bar{a}_t, \bar{y}_{t-1}) \text{d}y_t \\
= &\hspace{0.5em} \int_{\mathcal{L}_t \cap \mathcal{A}_t} (s_t(y_t; \bar{a}_T, \bar{y}_{t-1})-1)(f_{T, t}(y_t, a_T,\ldots,a_{t+1}; \bar{a}_t, \bar{y}_{t-1})-b_t(\bar{a}_T, \bar{y}_{t-1})) p_t(y_t|\bar{a}_t, \bar{y}_{t-1}) \text{d}y_t + \\
&\hspace{0.5em} \int_{\mathcal{U}_t \cap \mathcal{A}_t} (s_t(y_t; \bar{a}_T, \bar{y}_{t-1})-1)(f_{T, t}(y_t, a_T,\ldots,a_{t+1}; \bar{a}_t, \bar{y}_{t-1})-b_t(\bar{a}_T, \bar{y}_{t-1})) p_t(y_t|\bar{a}_t, \bar{y}_{t-1}) \text{d}y_t + \\
&\hspace{0.5em} \int_{(\mathcal{L}_t \cup \mathcal{U}_t) \backslash \mathcal{A}_t} (s_t(y_t; \bar{a}_T, \bar{y}_{t-1})-1)(f_{T, t}(y_t, a_T,\ldots,a_{t+1}; \bar{a}_t, \bar{y}_{t-1})-b_t(\bar{a}_T, \bar{y}_{t-1})) p_t(y_t|\bar{a}_t, \bar{y}_{t-1}) \text{d}y_t \\
\leq &\hspace{0.5em} \int_{\mathcal{L}_t \cap \mathcal{A}_t} (s_t(y_t; \bar{a}_T, \bar{y}_{t-1})-1)(f_{T, t}(y_t, a_T,\ldots,a_{t+1}; \bar{a}_t, \bar{y}_{t-1})-b_t(\bar{a}_T, \bar{y}_{t-1})) p_t(y_t|\bar{a}_t, \bar{y}_{t-1}) \text{d}y_t + \\
&\hspace{0.5em} \int_{\mathcal{U}_t \cap \mathcal{A}_t} (s_t(y_t; \bar{a}_T, \bar{y}_{t-1})-1)(f_{T, t}(y_t, a_T,\ldots,a_{t+1}; \bar{a}_t, \bar{y}_{t-1})-b_t(\bar{a}_T, \bar{y}_{t-1})) p_t(y_t|\bar{a}_t, \bar{y}_{t-1}) \text{d}y_t \\
\leq &\hspace{0.5em} -m\int_{\mathcal{L}_t \cap \mathcal{A}_t} (1 - s_t(y_t; \bar{a}_T, \bar{y}_{t-1})) p_t(y_t|\bar{a}_t, \bar{y}_{t-1}) \text{d}y_t \\
&\hspace{0.5em} - m\int_{\mathcal{U}_t \cap \mathcal{A}_t} (s_t(y_t; \bar{a}_T, \bar{y}_{t-1}) - 1) p_t(y_t|\bar{a}_t, \bar{y}_{t-1}) \text{d}y_t \\
= &\hspace{0.5em} -m\int_{\mathcal{A}_t} |1 - s_t(y_t; \bar{a}_T, \bar{y}_{t-1})|p_t(y_t|\bar{a}_t, \bar{y}_{t-1}) \text{d}y_t\\
< &\hspace{0.5em} 0.
\end{split}
\end{equation}

By applying conditions (\ref{prop:eq1}) and (\ref{prop:eq2}), we have for $t = 1,..., T-1$,
\begin{equation}
\label{prop:eq3}
\begin{split}
&\hspace{0.5em}\int s_t(y_t; \bar{a}_T, \bar{y}_{t-1})f_{T, t}(y_t, a_T,\ldots,a_{t+1}; \bar{a}_t, \bar{y}_{t-1})p_t(y_t|\bar{a}_t, \bar{y}_{t-1}) \text{d}y_t \\
= &\hspace{0.5em}\int_{\mathcal{N}_t} s_t(y_t; \bar{a}_T, \bar{y}_{t-1})f_{T, t}(y_t, a_T,\ldots,a_{t+1}; \bar{a}_t, \bar{y}_{t-1})p_t(y_t|\bar{a}_t, \bar{y}_{t-1}) \text{d}y_t+ \\
&\hspace{0.5em}\int_{\mathcal{L}_t} s_t(y_t; \bar{a}_T, \bar{y}_{t-1})f_{T, t}(y_t, a_T,\ldots,a_{t+1}; \bar{a}_t, \bar{y}_{t-1})p_t(y_t|\bar{a}_t, \bar{y}_{t-1}) \text{d}y_t+ \\
&\hspace{0.5em}\int_{\mathcal{U}_t} s_t(y_t; \bar{a}_T, \bar{y}_{t-1})f_{T, t}(y_t, a_T,\ldots,a_{t+1}; \bar{a}_t, \bar{y}_{t-1})p_t(y_t|\bar{a}_t, \bar{y}_{t-1}) \text{d}y_t \\
= &\hspace{0.5em}\int_{\mathcal{N}_t} f_{T, t}(y_t, a_T,\ldots,a_{t+1}; \bar{a}_t, \bar{y}_{t-1})p_t(y_t|\bar{a}_t, \bar{y}_{t-1}) \text{d}y_t+ \\
&\hspace{0.5em}\int_{\mathcal{L}_t} f_{T, t}(y_t, a_T,\ldots,a_{t+1}; \bar{a}_t, \bar{y}_{t-1})p_t(y_t|\bar{a}_t, \bar{y}_{t-1}) \text{d}y_t+ \\
&\hspace{0.5em}\int_{\mathcal{U}_t} f_{T, t}(y_t, a_T,\ldots,a_{t+1}; \bar{a}_t, \bar{y}_{t-1})p_t(y_t|\bar{a}_t, \bar{y}_{t-1}) \text{d}y_t+ \\
&\hspace{0.5em}\int (s_t(y_t; \bar{a}_T, \bar{y}_{t-1})-1)(f_{T, t}(y_t, a_T,\ldots,a_{t+1}; \bar{a}_t, \bar{y}_{t-1})-b_t(\bar{a}_T, \bar{y}_{t-1})) p_t(y_t|\bar{a}_t, \bar{y}_{t-1}) \text{d}y_t+\\
&\hspace{0.5em}\int b_t(\bar{a}_T, \bar{y}_{t-1})(s_t(y_t; \bar{a}_T, \bar{y}_{t-1}) - 1) p_t(y_t|\bar{a}_t, \bar{y}_{t-1}) \text{d}y_t\\
< &\hspace{0.5em}\int_{\mathcal{N}_t} f_{T, t}(y_t, a_T,\ldots,a_{t+1}; \bar{a}_t, \bar{y}_{t-1})p_t(y_t|\bar{a}_t, \bar{y}_{t-1}) \text{d}y_t+ \\
&\hspace{0.5em}\int_{\mathcal{L}_t} f_{T, t}(y_t, a_T,\ldots,a_{t+1}; \bar{a}_t, \bar{y}_{t-1})p_t(y_t|\bar{a}_t, \bar{y}_{t-1}) \text{d}y_t+ \\
&\hspace{0.5em}\int_{\mathcal{U}_t} f_{T, t}(y_t, a_T,\ldots,a_{t+1}; \bar{a}_t, \bar{y}_{t-1})p_t(y_t|\bar{a}_t, \bar{y}_{t-1}) \text{d}y_t \\
= &\hspace{0.5em}\int f_{T, t}(y_t, a_T,\ldots,a_{t+1}; \bar{a}_t, \bar{y}_{t-1}) p_t(y_t|\bar{a}_t, \bar{y}_{t-1}) \text{d}y_t.
\end{split}
\end{equation}
Iteratively apply the inequality (\ref{prop:eq3}) from $t=T-1$ to $t=1$. It follows that
\begin{equation*}
    \begin{split}
        \E[Y_T|\bar{a}_T] 
        &= \idotsint \E[Y_T|\bar{A}_T=\bar{a}_T, \bar{Y}_{T-1}=\bar{y}_{T-1}]\prod_{l=1}^{T-1} p_l(y_{l}|\bar{a}_{T}, \bar{y}_{l-1}) \text{d}y_l \\
        &= \idotsint \E[Y_T|\bar{A}_T=\bar{a}_T, \bar{Y}_{T-1}=\bar{y}_{T-1}]\prod_{l=1}^{T-1} s_{l}(y_l; \bar{a}_T, \bar{y}_{l-1}) p_l(y_{l}|\bar{a}_{l}, \bar{y}_{l-1}) \text{d}y_l \\
        &= \idotsint f_{T, {T-1}}(y_{T-1}, a_T; \bar{a}_{T-1}, \bar{y}_{T-2}) \prod_{l=1}^{T-1} s_{l}(y_l; \bar{a}_T, \bar{y}_{l-1}) p_l(y_{l}|\bar{a}_{l}, \bar{y}_{l-1}) \text{d}y_l \\
        &< \idotsint f_{T, {T-1}}(y_{T-1}, a_T; \bar{a}_{T-1}, \bar{y}_{T-2}) p_{T-1}(y_{T-1}|\bar{a}_{T-1}, \bar{y}_{T-2}) \text{d}y_{T-1}\\
        &\hspace{5em} \prod_{l=1}^{T-2} s_{l}(y_l; \bar{a}_T, \bar{y}_{l-1}) p_l(y_{l}|\bar{a}_{l}, \bar{y}_{l-1}) \text{d}y_l \\
        &= \idotsint f_{T, {T-2}}(y_{T-2}, a_T,a_{T-1}; \bar{a}_{T-2}, \bar{y}_{T-3}) \prod_{l=1}^{T-2} s_{l}(y_l; \bar{a}_T, \bar{y}_{l-1}) p_l(y_{l}|\bar{a}_{l}, \bar{y}_{l-1}) \text{d}y_l \\
        &< \ldots \\
        &= \int f_{T, 1}(y_1, a_T,\ldots,a_{2}; a_1, y_0) s_{1}(y_1; \bar{a}_T, y_{0}) p_1(y_1|a_{1}, y_0)\text{d}y_1 \\
        &< \int f_{T, 1}(y_1, a_T,\ldots,a_{2}; a_1, y_0) p_1(y_1|a_{1}, y_0)\text{d}y_1 \\
        &= \E[Y_T^{\bar{a}_T}].
    \end{split}
\end{equation*}

\end{proof}

\subsection*{Proof for Example}

\begin{proof}

At each $t = 1,..., T-1$, the adaptive propensity score ratio is
\begin{equation*}
s_t(y_t; a_T,\ldots,a_{t+1}, \bar{a}_t, \bar{y}_{t-1}) = \frac{p(a_T,\ldots a_{t+1}|\bar{a}_t, \bar{y}_{t})}{p(a_T, \ldots, a_{t+1}|\bar{a}_t, \bar{y}_{t-1})} = \frac{\Pi_{s=1}^{T-t}p(a_{t+s}|\bar{a}_{t-1+s},\bar{y}_{t})}{p(a_T, \ldots, a_{t+1}|\bar{a}_t, \bar{y}_{t-1})}.
\end{equation*}

The adaptive propensity score is well defined when $p(a_T, \ldots, a_{t+1}|\bar{a}_t, \bar{y}_{t-1}) > 0$. In other words, $y_s \leq \tilde{y}$ for each $s = 1, \ldots, t$.

When $y_t > \tilde{y}$, we have $p(a_{t+1}|\bar{a}_{t},\bar{y}_{t}) = 0$ so $s_t(y_t; a_T,\ldots,a_{t+1}, \bar{a}_t, \bar{y}_{t-1}) = 0$.

When $y_t \leq \tilde{y}$, $p(a_{t+s}|\bar{a}_{t-1+s},\bar{y}_{t})$ is monotonically decreasing with respect to $y_t$ for each $s = 1, \ldots, T-t$ due to the property of a monotonically increasing disease outcome process.

Therefore, $s_t(y_t; a_T,\ldots,a_{t+1}, \bar{a}_t, \bar{y}_{t-1})$ is monotonically decreasing with respect to $y_t$. When the adaptive propensity score ratio is not almost surely constant, the downweighted and upweighted adaptations are
\begin{equation*}
\mathcal{L}_t(a_T,\ldots,a_{t+1}; \bar{a}_{t}, \bar{y}_{t-1}) \subseteq (y_L, N] \text{ and } \mathcal{U}_t(a_T,\ldots,a_{t+1}; \bar{a}_{t}, \bar{y}_{t-1}) \subseteq (y_{t-1}, y_U),
\end{equation*}
where $y_{t-1} \leq y_U \leq y_L \leq \tilde{y}$ and $N$ is the size of the population. By invoking the property of the monotonically increasing disease outcome process again, for any $\bar{y}_{t-1} \in \{\bar{y}_{t-1}: p(a_T^*,\ldots a_{t+1}^*|\bar{a}_t, \bar{y}_{t-1}) > 0 \}$, we have
\begin{equation*}
\begin{split}
        &\inf_{y_t^L \in \mathcal{L}_t(a_T^*,\ldots,a_{t+1}^*; \bar{a}_t, \bar{y}_{t-1})} f_{T, t}(y_t^L, a_T^*,\ldots,a_{t+1}^*; \bar{a}_t, \bar{y}_{t-1})\\ 
        \geq &\sup_{y_t^U \in \mathcal{U}_t(a_T^*,\ldots,a_{t+1}^*; \bar{a}_t, \bar{y}_{t-1})} f_{T, t}(y_t^U, a_T^*,\ldots,a_{t+1}^*; \bar{a}_t, \bar{y}_{t-1}),
\end{split}
\end{equation*}

It only requires the weak regular condition that outcomes from the monotonically increasing process by $\mathcal{L}_t$ and $\mathcal{U}_t$ are separated on a nonempty subset where the L-1 norm of the absolute difference between $s_t(y_t; a_T,\ldots,a_{t+1}, \bar{a}_t, \bar{y}_{t-1})$ and $1$ is nonzero. The intervention $\bar{a}_T = (a_1, \ldots, a_T) = (0, \ldots, 0)$ is thus opportunistic at each $t$. By Theorem \ref{thrm:TVCB}, the time-varying confounding bias is negative.

\end{proof}

\bibliography{main_arxiv_submission}
\bibliographystyle{abbrvnat}

\end{document}